**Indigenous rights, peoples, and space exploration: A response to the Canadian Space Agency (CSA) Consulting Canadians on a framework for future space exploration activities**

Hilding Neilson[1] and Elena E. Ćirković[2]
1. David A. Dunlap Department of Astronomy & Astrophysics, University of Toronto. hilding.neilson@utoronto.ca
2. HELSUS, University of Helsinki

**Abstract**

Canada is beginning to plan its next chapter of space exploration that includes sending humans back to the Moon and onwards to Mars.  This includes understanding humanities place in space and who will benefit from our exploration.  As part of this plan the Canadian Space Agency (CSA) placed a call for consultations.  In response, we presented comments urging the CSA to be inclusive of Indigenous peoples in the planning as well as to be inclusive of Indigenous rights and worldview in the future of space exploration.  In particular, we explore the questions of how Outer Space Laws intersect with treaties between Indigenous Nations and the Crown in what is today Canada, how the current narratives of space exploration parallel the historic narratives of colonization that negatively impact Indigenous peoples, and how the future of commercial exploitation of outer space acts to further colonization.

**Introduction**

Canada's position of support and leadership in space exploration has a positive and impressive history. From the development of the Canada Arm and the participation in work on the International Space Station (ISS) to the new scientific contributions with respect to lunar and Martian exploration, Canada has many reasons to be proud.  However, it is worth noting that Canada's role in space exploration has traditionally neglected to include Indigenous peoples, Indigenous knowledges, and Indigenous rights.  In general, the history of Canadian participation in space exploration did not have substantial and direct impact on Indigenous peoples' rights in Canada. With accelerating technological developments in the past twenty years, space has become more accessible for humans.  With these transformations, the current and proposed future of space exploration has the potential to negatively impact Indigenous peoples across Canada. Without consultation with multiple knowledges of multicultural and multinational Canada, future space activities might contribute to the ongoing culture of colonization.

In this submission we present arguments for the ethical and requirements for the Canadian Space Agency (CSA) to consult with and to be inclusive of Indigenous rights and concerns as Canada moves to support the Artemis Accords. The authors come to this work from two perspectives: the first being a Mi'kmaw astronomer who grew up in Newfoundland and is a status member of the Qalipu Nation, and second, a legal scholar of minority background.  Thereby we stress that our contribution is an opinion and has no intent to speak for Indigenous peoples in general, and/or any Indigenous-led organization in Canada, or any particular group or community in Canada. Please note that we will be using the terms "Indigenous", and "Aboriginal" interchangeably as we engage with the language of domestic (Canadian) and international documents, publications, institutions and relevant regulatory and/or administrative bodies. The terms Indigenous and "Aboriginal" refers to the three different categories of Indigenous peoples in Canada - First Nation, Inuit, and Métis.

In this work we reflect upon the CSA's obligation to consult Indigenous peoples in Canada via two lenses:

1) Where does Outer Space Law intersect with the modern and historic treaties between the First Nations and Canada (Crown)? Do these treaties include the skies and outer space?
2) Considering its status as an international (and bilateral) agreement, where the Artemis Accords trigger the application of the United Nations Declaration on the Rights of Indigenous Peoples, GA Res 61/295, UNGAOR, 61st Sess, UN Doc A/RES/61/295 (2007), Assuming that the Artemis Accords might, and in the situations where they do, trigger any responsibilities and obligations of Canada under UNDRIP and its domestic laws, to consult the First Nations, what are the CSA's and Canada's obligations to First Nation, Inuit, and Métis communities and Nations?

We engage with these two points considering the following:
1) That the questions of Indigenous rights and title in Canada, including the treaty rights, have significant impacts on how Canada consults with the First Nations and other communities and nations in Canada and pursues accordingly the ongoing and future space exploration;
2) That these questions also require a revisiting of the allegedly prevailing narrative as proposed by some scholars and members of the global outer space sector, generally speaking, which treats space exploration as an analogy of the colonization of the Americas (e.g., Zubrin 2011).

First, we briefly consider the status of Indigenous peoples' rights in Canada. Second, we address the narrative of space exploration. In the third section we address the ongoing issue of megaconstellations of satellites, and finally, the fourth section addresses human activities on Moon and Mars.

**Brief consideration of Indigenous Rights in Canada**

Canada's obligations to Indigenous peoples under the Canadian Constitution cannot be superseded or undermined by commitments under a bilateral agreement such as the Artemis Accords. These legal obligations include those recognized and affirmed by Section 35 of the *Constitution Act*, 1982, and those set out in self-government agreements.

We recognize that, in 1985, the Supreme Court of Canada (SCC) concluded that treaties between Indigenous peoples and the Crown were not international treaties but were *sui generis* treaties (Simon v The Queen, [1985] 2 SCR 387 at para 33). However, it is worth considering that, "[f]or many Indigenous peoples, treaties concluded with European powers…are, above all, treaties of peace and friendship, destined to organize coexistence in — not their exclusion from — the same territory and not to regulate restrictively their lives…under the overall jurisdiction of non-Indigenous authorities" (UNESC, 1999 at para. 117).

While the United Nations, in documents including the UNDRIP has recognized the potentially international character of Indigenous Crown treaties (UNDRIP Preamble, art 37(1) [UN Declaration], we recognize that Canadian law has yet to consider this international recognition in domestic law. Nevertheless, as Henderson argues "any Crown authority over First Nations is limited to the actual scope of their treaty delegations. If no authority or power is delegated to the Crown, this power must be interpreted as reserved to First Nations, respectively, and is protected by prerogative rights and the common law since neither can extinguish a foreign legal system." (Henderson, 1994, at 268). We wish to stress that there are plural and ongoing discussions on the status of Aboriginal title in Canada, as well as treaty obligations. It is beyond the scope of our comment to address the extensive international and domestic jurisprudence on the topic. However, we stress the existence of Crown's Fiduciary Duty to

Aboriginal People as an aspect of various activities, including Canada's activities in outer space (See, Annex I). Indeed, "The doctrine of Aboriginal rights exists… because of one simple fact: when Europeans arrived in North America, Aboriginal peoples were already here, living in communities on the land, and participating in distinctive cultures, as they had done for centuries.  It is this fact, and this fact above all others, which separates Aboriginal peoples from all other minority groups in Canadian society and which mandates their special legal status." ( Chief Justice Lamer in *R. v. Van der Peet*, para 30).

**The Narratives of Space Exploration**

Human space exploration has been part of the social and political conscious since Dr. Werner von Braun published his seminal guidebook for a human settlement of Mars (von Braun & White 1953). The possibility of human settlement in space, has also prompted the consideration of  parallels between the narratives of space exploration and colonization. These parallels include simple phrases such as "colonization" and "frontier" but are related in the concepts of the Doctrine of Discovery, Manifest Destiny and Terra Nullius. These parallels are well understood in popular culture and science fiction (e.g., King, 2017). Indeed, as and further, since the beginning of the space race, the drafters of the 1967 Outer Space Treaty (OST), to which Canada is a party, agreed that "colonization" or the possibility of a "land grab" in outer space, was to be avoided. For this reason, outer space became a "global commons".

The first issue is our choice of words with respect to space exploration.  The words that are typically chosen, such as colonization, is exclusive to peoples who have a lived experience of negative impacts of colonization. In Canada, the impacts of this colonization includes recent land defense protests in Wet'su'weten, southern Ontario; the ongoing and historical removal of Indigenous children from their families that is traced back to the 60's scoop; the intergenerational traumas of Residential School; the government-sanctioned starvation and abuse of Indigenous peoples in the early history of Canada as a nation state; the bounties offered for scalping First Nation peoples, and so on (e.g., Diabo 2020, King 2013, Palmater 2011). These histories have been well documented in scholarship on the Canada's history in relation of the First Nations and is beyond the scope of our commentary. However, we stress that focusing on the ideas of "frontiers" or "colonization", when discussing the human presence in outer space, echoes the remnants of this history and excludes Indigenous peoples from participating in the goal of space exploration.

One counterargument to such concerns could be that human exploration in space will not impact "life" as there is no "life" as we know it, on the Moon and that most likely, there is no life on Mars.
Responses to this are as follows:
1) Definition of "life" is rooted in dominant ontologies, and a particular understanding of physical entities that have biological processes and are distinguished from inorganic matter. However, many Indigenous peoples in Canada and around the world have different perspectives of life and being.  For many peoples there is a kinship relation between humans and nature, humans and the land, and humans and celestial objects such as the planets, Moon, stars, and the Sun as well as the Universe.  This is an axiom of relationality that places Indigenous peoples as equal and to above any other element in nature.
2) The argument that outer space and space objects are "empty" are a space-based transposition of the *terra nullius* (Latin: nobody's land)  or *res nullius* (nobody's thing) principles in international law (or the early law among the nations).  Terra Nullius and the Doctrine of Discovery are principles developed by colonizing European Nations, and philosophies and laws, which declared territories as empty, and avoid recognition of the existence of Indigenous peoples. For instance, both, the idea that Indigenous peoples' land was empty land, and the Doctrine of Discovery, are exemplified in the 1494 *Treaty of Tordesillas*, which declared that only non-Christian lands could be colonized under the Doctrine of Discovery. In the nineteenth century, the famous 1823 US Supreme Court case *Johnson v. M'Intosh* 1823 affirmed the doctrine of discovery as part of international law. As Chief Justice John Marshall noted in his opinion as follows:

> On the discovery of this immense continent, the great nations of Europe ... as they were all in pursuit of nearly the same object, it was necessary, in order to avoid conflicting settlements, and consequent war with each other, to establish a principle which all should acknowledge as the law by which the right of acquisition, which they all asserted, should be regulated as between themselves. This principle was that discovery gave title to the government by whose subjects, or by whose authority, it was made, against all other European governments, which title might be consummated by possession. ... The history of America, from its discovery to the present day, proves, we think, the universal recognition of these principles.

These are examples of how doctrine of discovery was used as a legal instrument. However, in 2021, and in the context of current general international law, colonialism is no longer legal (McWhinney 2008).

3) These policies allowed European nations to claim territory through colonialism and erase the connections between Indigenous peoples and the lands on which they lived on by declaring those Indigenous peoples as not civilized and often as not 'human'-and therefore, not as subjects of law. While we have no reason to expect there are any humans nor any living creature that western science would recognize as "intelligent" living currently on the Moon or Mars, the concept of Terra Nullius, or Lunar Nullius or Martian Nullius impacts all living beings in the respective areas regardless of perceived intelligence. Terra Nullius impacted humans in what came to be the sovereign Canada, as well as bison, wolves, bears, cod, salmon, great auk, and so on. It is arguable that Terra Nullius negatively impacted the rights of the land, through dam building, water and air pollution, and so on. We don't know what the impact of Martian Nullius is, or will be, and our narratives and discussions must be inclusive of these issues from the perspective of Indigenous knowledges and ontologies, in Canada.

**Megaconstellations of Satellites is Colonization**

The launch of Starlink by SpaceX has had a dramatic and damaging impact on research in astronomy and astrophysics (Clery 2020, Kocifaj 2021). These satellites have added to the amount of light pollution and future satellite constellations could have far greater impact depending on the legal requirements and the purpose of those satellites.

Hamacher et al (2020) presented a compelling argument that light pollution is a form of cultural genocide (please note that in the context of the Final Report of the Truth & Reconciliation Commission we will use the term Indigenous erasure instead). In their article, the authors noted that a significant amount of Indigenous knowledge is based on star lore and observations of the sky. Those observations are connected to Indigenous stories about the land and nature - for some peoples the sky is a reflection of the land (Cajete 2000). Those observations, however, are based on a dark night sky without substantive light pollution. As such, light pollution acts to disconnect Indigenous peoples from the land they live, and as such, is a form of erasure. In the same vein, we argue that constellations of satellites are also a form of colonization, especially those that are bright enough to be visible from the ground. If light pollution results in an erasure of knowledges, then megaconstellations of satellites would also constitute an attempt to rewrite that knowledge.

There is a second issue that the CSA should consider with respect to space exploration and the impact of new satellites. That issue is at what height do treaties and agreements with Indigenous peoples, end? It is understood that treaties have impact on Indigenous rights and responsibilities with respect to mining, water resources, hunting, etc. but Indigenous communities should be consulted with the impacts on the skies above. This is especially true for satellites that contribute to light pollution, but also satellites that are designed to offer services to communities (such as wireless internet), satellites designed for ground-based or remote imaging such as mapping satellites and LIDAR imaging. The CSA has an obligation to consult with Indigenous communities and Indigenous-led organizations with respect to the legalities of how satellites that impact communities operate.

**Preserving the Moon and Mars**

One of the key elements of the Artemis Accords is the commitment to preserve Outer Space Heritage (Section 9). On the other hand, Sections 3 and 10 of the Accords are designed to allow countries to peacefully exploit the Moon and Mars. However, these accords presuppose that Space Heritage refers to only landing sites and rovers. This definition ignores Indigenous people's perspectives and elements of space heritage for Indigenous cultures.

It is also notable that the accords allow for exploitation by humanity for industry such as mining. This idea implies that nation-states on Earth have the right to exploit the Moon and Mars for their own purposes and those rights supersede the principle that the Moon and Mars might have its own rights as viewed from Indigenous perspectives. In Aotearoa (New Zealand), the settler government recognized that the Whanganui River is a living entity that belonged to no one, hence has its own rights as a living entity. This means that the river cannot be exploited by humans. Because the river cannot express its own interests in ways that humans can interpret a committee was appointed to act as guardian for the river. That committee includes local Maori representatives.

The importance of the Moon and Mars as part of the cultures and knowledge systems of Indigenous peoples from around the world and that part for many peoples is one of relation (Cajete, 2000). In the situation of relationality, the Moon and Mars and other Solar Systems objects have their own rights to exist and be. Those rights are not necessarily incongruent with exploration and mining. However, they do require a significant reconsideration of what constitutes a human right to interact with the Moon and Mars. For instance, the environmental impact on the Earth is significant but the Earth could heal from most mining events given enough time. It is not likely the Moon would recover over the age of the Universe because of the lack dynamic mechanisms found on the Earth. We have an ethical duty to consider the rights of the Moon and Mars from environmental and Indigenous perspectives to better share the benefits and sustainability of space exploration. Those rights should be represented by Indigenous peoples as well as traditional nation-state governance

**Annex I**

Brief summaries of the court cases on the Aboriginal rights and title:

**Calder v British Columbia (AG) [1973] SCR 313, [1973] 4 WWR 1**
Calder decision recognizes Aboriginal title.

**Guerin v. The Queen, [1984] 2 S.C.R. 335 Date : 1984-11-01**
Guerin decision established Crown's fiduciary duty to protect Aboriginal title.
The Supreme Court ruled that the federal government had a "fiduciary responsibility" for Indians and lands reserved for Indians - that is, a responsibility to safeguard their interests. This duty placed the government under a legal obligation to manage Aboriginal lands as a prudent person would when dealing with his/her own property.

**R. *v.* Sparrow, [1990] 1 S.C.R. 1075**
The 1990 Supreme Court Decision in R. v. Sparrow was the first Supreme Court of Canada decision which applied Section 35, of the Constitution Act, 1982 which states "The existing Aboriginal and treaty rights of the Aboriginal peoples of Canada are hereby recognized and affirmed".

**Delgamuukw *v.* British Columbia, [1997] 3 S.C.R. 1010**
Delgamuukw decision confirms Aboriginal title exists.
The three commonly called Delgamuuk cases are a critical part of the constitutional evaluation of Aboriginal rights and title for British Columbia and all of Canada. This case is of particular importance as it recognized the importance of oral tradition as a defining aspect of a First Nations culture and accepted the oral history presented at trial. This means that the oral traditions can be used as evidence to determine Aboriginal rights and title, and with recognition of what is of cultural significance to them as part of their rights and title.

**Haida Nation *v.* British Columbia (Minister of Forests), [2004] 3 S.C.R. 511, 2004 SCC 73**
Haida decision established the duty to consult & accommodate.
This was a unanimous decision of the Supreme Court of Canada as it set out the basic principles applicable to the duty to consult.

**Tsilhqot'in Nation v. British Columbia, 2014 SCC 44, [2014] 2 S.C.R. 256. Date: 20140626. Docket: 34986**
Tsilhqot'in took the test to prove title and applied it. The Supreme Court of Canada awarded the land to the Tsilhqot'in Nation.

**REFERENCES**

**Books and Articles**

King, T. (2017). The inconvenient Indian illustrated: A curious account of native people in North America. Doubleday Canada.

King, S. (2013). *Fishing in contested waters: Place & community in burnt church/Esgenoopetitj*. University of Toronto Press.

Kocifaj, M., Kundracik, F., Barentine, J.C., & Bará, S. (2021) *The proliferation of space objects is a rapidly increasing source of artificial night sky brightness*. MNRAS, in press.

Palmater, P. D. (2011). *Beyond blood: Rethinking Indigenous identity*. UBC Press.

Von Braun, W., & White, H. J. (1953). *The Mars Project*. University of Illinois Press.

Zubrin, R. (2011). *Case for Mars*. Simon and Schuster.

**Canadian Supreme Court Cases**

Simon v The Queen, [1985] 2 SCR 387 at para 33

**US Supreme Court Cases**

"Johnson v. M'Intosh", 21 U.S. 543, 5 L.Ed. 681, 8 Wheat. 543 (1823)

**UN Documents**

McWhinney, E. (2008) Declaration on the Granting of Independence to Colonial Countries and Peoples available at: https://legal.un.org/avl/pdf/ha/dicc/dicc_e.pdf

United Nations Declaration on the Rights of Indigenous Peoples, GA Res 61/295, UNGAOR, 61st Sess, UN Doc A/RES/61/295 (2007)

UN Economic and Security Council, Sub-Commission on Prevention of Discrimination and Protection of Minorities, Human Rights of Indigenous Peoples: Study on treaties, agreements and other constructive arrangements between States and Indigenous populations, Final report by Miguel Alfonso Martínez, Special Rapporteur, 51st Sess, UN Doc E/ CN.4/Sub.2/1999/20 (1999) at para 117).